\documentclass[12pt]{article}
\pdfoutput=1

\usepackage[nosort]{cite}
\usepackage[hyperref,bulletsep]{collect}

\usepackage{graphicx}
\usepackage{bbm}
\usepackage{amsmath,amssymb}
\usepackage{subfigure}
\usepackage{verbatim}

\makeatletter
\let\old@startsection=\@startsection
\let\oldl@section=\l@section
\renewcommand{\@startsection}[6]{\old@startsection{#1}{#2}{#3}{#4}{#5}{#6\mathversion{bold}}}
\renewcommand{\l@section}[2]{\oldl@section{\mathversion{bold}#1}{#2}}
\makeatother

\makeatletter
\let\old@makecaption=\@makecaption
\def\@makecaption{\small\old@makecaption}
\makeatother

\makeatletter \@addtoreset{equation}{section} \makeatother

\let\oldPhi=\Phi
\let\oldPsi=\Psi
\let\oldGamma=\Gamma
\let\oldDelta=\Delta
\let\oldSigma=\Sigma
\let\oldTheta=\Theta
\let\oldPi=\Pi
\let\oldUpsilon=\Upsilon
\renewcommand{\Phi}{\mathnormal{\oldPhi}}
\renewcommand{\Psi}{\mathnormal{\oldPsi}}
\renewcommand{\Gamma}{\mathnormal{\oldGamma}}
\renewcommand{\Sigma}{\mathnormal{\oldSigma}}
\renewcommand{\Delta}{\mathnormal{\oldDelta}}
\renewcommand{\Theta}{\mathnormal{\oldTheta}}
\renewcommand{\Pi}{\mathnormal{\oldPi}}
\renewcommand{\Upsilon}{\mathnormal{\oldUpsilon}}

\newcommand{\Tr}{\mathop{\mathrm{Tr}}}

\renewcommand{\Im}{\mathop{\mathrm{Im}}}

\newcommand{\sech}{\mathop{\mathrm{sech}}}

\newcommand{\eps}{\epsilon}

\newcommand{\order}{\mathcal{O}}

\newcommand{\A}{\textbf{A}}
\newcommand{\B}{\textbf{B}}
\newcommand{\C}{\mathcal{C}}

\newcommand{\superN}{\mathcal{N}}
\newcommand{\NN}{\mathcal{N}}
\newcommand{\SM}{\textbf{\textrm{S}}}

\renewcommand{\d}{\partial}

\newcommand{\ve}{{\mathcal E}}

\newcommand{\up}{\,\uparrow\,}
\newcommand{\down}{\,\downarrow\,}

\def\Xint#1{\mathchoice
  {\XXint\displaystyle\textstyle{#1}}%
  {\XXint\textstyle\scriptstyle{#1}}%
  {\XXint\scriptstyle\scriptscriptstyle{#1}}%
  {\XXint\scriptscriptstyle\scriptscriptstyle{#1}}%
  \!\int}
\def\XXint#1#2#3{{\setbox0=\hbox{$#1{#2#3}{\int}$}
    \vcenter{\hbox{$#2#3$}}\kern-.5\wd0}}
\def\pint{\Xint-}

\newcommand{\sfrac}[2]{{\textstyle\frac{#1}{#2}}}
\newcommand{\half}{\sfrac{1}{2}}

\newcommand{\Half}{\frac{1}{2}}

\newcommand{\lrbrk}[1]{\left(#1\right)}

\newcommand{\Biggsbrk}[1]{\Biggl[#1\Biggr]}

\newcommand{\be}{\begin{equation}}
\newcommand{\ee}{\end{equation}}
\newcommand{\nn}{\nonumber}

\makeatletter
\def\mr@ignsp#1 {\ifx\:#1\@empty\else #1\expandafter\mr@ignsp\fi}%
\newcommand{\multiref}[1]{\begingroup
\xdef\mr@no@sparg{\expandafter\mr@ignsp#1 \: }%
\def\mr@comma{}%
\@for\mr@refs:=\mr@no@sparg\do{\mr@comma\def\mr@comma{,}\ref{\mr@refs}}%
\endgroup}
\makeatother

\newcommand{\hypref}[2]{\ifx\href\asklfhas #2\else\href{#1}{#2}\fi}

\newcommand{\figref}[1]{figure~\multiref{#1}}
\renewcommand{\eqref}[1]{(\multiref{#1})}

\ifx\href\asklfhas\newcommand{\href}[2]{#2}\fi

\begin{document}
\overfullrule=0pt
\parskip=2pt
\parindent=12pt
\headheight=0in \headsep=0in \topmargin=0in \oddsidemargin=0in

\vspace{ -3cm} \thispagestyle{empty} \vspace{-1cm}

\begin{flushright} UUITP-20/07
\end{flushright}

\begin{center}
\vspace{1.01cm}
{\Large\bf
Finite size effects for giant magnons\\ on physical strings
\vspace{.3cm}
}

 \vspace{.5cm} {
 J.\,A.~Minahan and O.~Ohlsson~Sax\footnote{joseph.minahan, olof.ohlsson-sax ``AT'' teorfys.uu.se}}

 \vskip 0.3cm

{
\em Department of Physics and Astronomy\\
Division of Theoretical Physics\\
Box 803\\
751 08 Uppsala, Sweden}

\end{center}

\vskip1cm
\begin{abstract}
  Using finite gap methods, we find the leading order finite size
  corrections for an arbitrary number of giant magnons on physical
  strings, where the sum of the momenta is a multiple of $2\pi$.  Our  results are valid for the Hofman-Maldacena fundamental giant magnons as well as their dyonic generalizations.  The energy corrections turn out to be
  surprisingly simple, especially if all the magnons are fundamental, and at leading order are independent of the magnon
  flavors.  We also show how to use the Bethe ansatz to find finite size corrections for dyonic giant  magnons with  large $R$-charges.
 \end{abstract}
\newpage

\tableofcontents

\section{Introduction}

Integrability in $\superN = 4$ planar gauge theories \cite{Minahan:2002ve,Beisert:2003tq,Beisert:2003yb} has provided many deep insights into the AdS/CFT correspondence \cite{Maldacena:1997re,Witten:1998qj, Gubser:1998bc}.  Recent spectacular progress \cite{Beisert:2006ez,Beisert:2006ib} in finding expressions for infinite dimensional operators valid for all values of the coupling rely on the existence of an integrable \SM-matrix \cite{Staudacher:2004tk,Beisert:2005fw,Beisert:2005tm,Janik:2006dc,Arutyunov:2004vx,Hernandez:2006tk} for a spin chain with infinite extent and long range interactions. 

Still, there remain many unsolved problems.  Chief among these is the question of what happens for operators of finite size.  In this case, wrapping effects arise due to the long range nature of the interactions \cite{Serban:2004jf}, and appear to spoil the Bethe ansatz equations in \cite{Beisert:2005fw}.  Nonetheless, there has been some progress in this direction.  In \cite{Ambjorn:2005wa}, Ambjorn, Janik and Kristjansen gave a systematic analyis of wrapping effects using the Thermodynamic Bethe ansatz.  Their analysis showed that wrapping effects will first appear at the $L$ loop level for an operator which corresponds to a spin chain with $L$ sites, precisely as expected.  Further analysis was carried out by Kotikov \textit{et. al.}  \cite{Kotikov:2007cy} who argued that wrapping effects must be included, lest there be a breakdown in the BFKL constraints at 4 loop order.  There have also been some recent explicit computations of the four loop contribution to the Konishi operator anomalous dimension \cite{Fiamberti:2007rj,Keeler:2008ce}, where wrapping effects lead to a $\zeta(5)$ term, a factor not present in the asymptotic Bethe equations \footnote{While both \cite{Fiamberti:2007rj} and \cite{Keeler:2008ce} find a $\zeta(5)$ contribution, there is a discrepancy between the two results.}.  

On the string theory side, finite size effects were computed for spinning strings by Sch\"afer-Nameki, Zamaklar and Zarembo \cite{SchaferNameki:2006ey}.  Here they considered a large spin $S$ and $R$-charge $J$ and found that the asymptotic Bethe equations in \cite{Beisert:2005fw}, including the appropriate dressing factors in \cite{Arutyunov:2004vx} and \cite{Hernandez:2006tk} give the correct answer for the energy levels up to terms of order $\exp(-2\pi J/\sqrt{\lambda})$, where $\lambda$ is the 't Hooft coupling.

Another direction for investigating finite size effects is to compute the corrections for giant magnon \cite{Hofman:2006xt} energies.  On a chain where the $R$-charge $J$ is taken to infinity, the energy of a single fundamental magnon on the resulting infinite chain is \cite{Beisert:2005tm}
 \begin{equation}\
 \ve =\sqrt{1+16g^2\sin^2\sfrac{p}{2}}\,,
 \end{equation}
 where $g$ is related to the 't Hooft coupling by
 \begin{equation}
 g=\frac{\sqrt{\lambda}}{4\pi}\,,
 \end{equation}
 and $p$ is the world-sheet momentum of the magnon.
 In the classical world-sheet limit where $g \gg 1$, the corresponding giant magnon solution was found by Hoffman and Maldacena, where the classical result is 
 \begin{equation}\label{classical}
 \ve \approx 4g\sin\sfrac{p}{2}\,.
\end{equation}
One can also consider $Q$ bound magnons \cite{Dorey:2006dq}, where the resulting energy is 
\begin{equation}\label{nonzeroQ}
 \ve =\sqrt{Q^2+16g^2\sin^2\sfrac{p}{2}}\,.
 \end{equation}
 The quantity $Q$ is the value of a second $R$-charge, and if $Q \gg 1$, the result in \eqref{nonzeroQ} can be found classically, the so-called dyonic giant magnon \cite{Chen:2006gea}.  One can think of the classical result in \eqref{classical} as the limiting case of \eqref{nonzeroQ} when $Q=0$ \footnote{$Q=0$ is the classical charge of a fundamental giant magnon.  The actual charge is $Q=1$, which  is  a consequence of fermion zeromodes \cite{Hofman:2006xt,Minahan:2007gf}.}.
 
 One can then consider corrections to the energies of \eqref{classical} and \eqref{nonzeroQ} for finite but very large $J$, the criteria being that $J \gg g$ and $J \gg Q$.  Such a calculation was carried out by Arutyunov, Frolov and Zamaklar, using a light cone gauge-fixing on the string world-sheet, where they found that the leading correction in the $Q=0$ case is \cite{Arutyunov:2006gs}
 \be\label{AFZ}
\Delta\ve \simeq-\,16\, g\,\sin^3\sfrac{p}{2}\,e^{-2}\exp\left(-2\frac{J}{4g\sin\frac{p}{2}}\right)
\ee
(An implicit result was also given in \cite{Arutyunov:2006gs} for the  $Q\ne0$ dyonic case.)  The result in \eqref{AFZ} is for a single magnon. In an entirely different approach, Janik and {\L}ukowski derived the correction in \eqref{AFZ} using Luscher's method for finite size effects in a relativistic quantum field theory.  The approach of these authors required the full BHL/BES \SM-matrix, providing a further check of the dressing phase in \cite{Beisert:2006ib}. 

However, strictly speaking there cannot be only one magnon, since in this case the total momenta of the magnons is not a multiple of $2\pi$, which violates the Virasoro constraints.  From the point of view of the dual gauge theory, having a total momenta that is not a multiple of $2\pi$ violates the trace condition.   Hence, for a single magnon one would expect gauge dependence in the final answer.  Indeed in \cite{Arutyunov:2006gs} it was shown that the higher order correction terms were dependent on a gauge parameter.  
One way around the predicament is to put the theory on a $Z_n$ orbifold.  In this case the total momentum only needs to be a multiple of $2\pi/n$.  This was done in \cite{Astolfi:2007uz} where they confirmed the result in \eqref{AFZ}.  Or one could  just have $n$ copies of the magnons, so that one finds the limit of the GKP spinning string on $S^5$ \cite{Gubser:2002tv} or its multicusp generalization \cite{Kruczenski:2004wg}.

But another way to proceed is to simply add more magnons, such that the total momentum is $0$ mod $2\pi$. With the Virasoro constraints now satisfied the results should be completely gauge independent.  However, in this case the presence of the other magnons would be expected to modify the correction to each  magnon's energy.

In this paper we will explicitly compute the leading order corrections to the magnon energy in the presence of other magnons.  We also include the possibility of bound magnons in the analysis.  Our main method employs the finite gap method in \cite{Kazakov:2004qf} for the string $\sigma$-model constrained to an $R\times S^3$ subspace.  $S^3$ has an $SU(2)\times SU(2)$ isometry and so our classical string configurations come with two independent $R$-charges.

Solutions of the finite gap equations will have $2M$ square root branch cuts, where $M$ is the number of giant magnons (some or all of these giant magnons could be dyonic).    Hence the general solutions will involve genus $2M-1$ hyperelliptic functions.  However, in the limit that $J\to\infty$ the hyperelliptic surface is highly singular with the branch cuts shrinking off to zero size.  In this limit it is possible to make a vast simplification and approximate the hyperelliptic functions with ordinary trigonometric functions.  One of our main results is that for $M$ fundamental giant magnons, the energy correction for each magnon is
\be\label{main}
\Delta\ve_j=-16\,g\,\sin^3\sfrac{p_j}{2}\prod_{k\ne j}^M\frac{\sin^2\frac{p_j+p_k}{4}}{\sin^2\frac{p_j-p_k}{4}}\,\exp(-2(J+\sum_k\ve_k)/\ve_j)\qquad 0\le p_k<2\pi
\ee
Note that the expression $J+\sum\ve_k$ is the total energy $E$  to leading order and can be thought of as the natural length of the chain since each magnon has a size equal to its energy.  Hence the natural suppression factor is $e^{-2E/\ve_j}$, not $e^{-2J/\ve_j}$.

We can also extend our results to nonzero values of $Q$, although the final expressions will not have quite the elegant form as in \eqref{main}.  Moreover, the exponential suppression term is modified to 
\be
e^{-2E H_j/\ve_j}\,,
\ee
where
\be
H_j=\frac{Q_j^2\sin^2\frac{p_j}{2}+16g^2\sin^4\frac{p_j}{2}}{Q_j^2+16g^2\sin^4\frac{p_j}{2}}\,.
\ee
It is now also possible to compare these results directly with the Bethe ansatz.  In the limit where all $Q_j \gg g$, the string result will approach that found for the one-loop gauge theory results.  In this case the problem reduces to finding the spectrum of the Heisenberg spin chain, where one can explicitly apply Bethe's ansatz.  The computation is significantly simpler but will still agree with the finite gap result.

In section 2 we show how to derive finite gap equations from the Heisenberg spin-chain and then give the modifications for the $R\times S^3$ $\sigma$-model, referring the reader to the references for the details of this latter part.  We then present the general configuration for a collection of giant magnons at large but finite $J$.  In section 3 we solve the gap equations to leading order for a generic giant magnon configuration.  We then go through some of the specific cases.  In section 4 we show how to find similar results for the Heisenberg spin chain and how these match onto the solutions in the previous sections.  In section 5 we discuss how to reproduce the results for a single set of bound magnons using TBA analysis, obtaining the same exponential term. In section 6 we present our conclusions and suggestions for further work.  

The results appearing here were first presented in \cite{Minahan:2007aa}.  As this paper was being typed, we received \cite{Hatsuda:2008gd} which considers the finite size corrections for a single dyonic giant magnon.

\section{From Heisenberg to finite gap}

In this section we give a short review for applying the Bethe ansatz to the one-loop anomalous dimension for single trace operators with large $R$-charge.   We show how to derive a set of integral equations and then generalize these equations to the finite gap solutions of the string sigma model on $R\times S^3$.  

The starting point is the chiral primary operator $\Tr Z^J$, whose dimension is protected by supersymmetry.  The field $Z$ is a complex adjoint scalar and the $R$-charge $J$ is a conserved global charge.  We can build other operators by inserting impurities between the $Z$-fields to give 
\be
\Tr[ Z\dots Z\chi_1 Z\dots Z\chi_2 Z\dots Z \chi_3 Z\dots Z]\ee
where the $\chi_i$ are one of eight boson or eight fermion fields.  For what follows we will restrict ourselves to the $SU(2)$ sector, so the operators we consider will be of the form
\begin{eqnarray}
&&\Tr [ Z\dots \,Z W Z\dots \,Z W Z\dots \,Z WW Z\dots]\nonumber\\
& \sim& \Tr[\up\dots\up\down\up\dots\up\down\up\dots\up\down\down\up\dots]\,,
\end{eqnarray}
where we have shown the map between operators of this type and a spin chain with up and down spins.
In \cite{Minahan:2002ve} it was argued that the one-loop anomalous dimensions for operators of this type are equivalent to the energies of the Heisenberg spin chain with Hamiltonian
\be\label{heisham}
H=L+2g^2\sum_{\ell=1}^L\left(\Half-2\,\vec S_\ell\cdot\vec S_{\ell+1}\right)
\ee
where $L$ is the total number of sites on the lattice.  If we assume that there are $J$ $Z$-fields and $Q$ $W$-fields then $L=J+Q$.  Each of the $R$-charges $J$ and $Q$ are conserved, and will be conserved to all orders in the perturbative expansion.  Note that we have included a constant term $L$ in the Hamiltonian so that the energies will match the full dimension up to the one-loop level.

The Heisenberg spin-chain is an integrable system and can be solved by finding a set of momenta for the impurities on the chain.  The impurities are the down spins, in other words the $W$-fields.  Each momentum satisfies a quantization condition 
\be
e^{ip_j(J+Q)}=\prod_{k\ne j}^Q S^{-1}(p_j,p_k)
\ee
where $S(p_j,p_k)$ is the \SM-matrix for impurities $j$ and $k$ and we write the number of lattice sites as a sum of the number of $Z$ and $W$ fields for later convenience.  It is convenient to express the momentum in terms of a rapidity variable $u_j$, where
\be\label{quantcond}
e^{ip_j}=\frac{u_j+i/2}{u_j-i/2}\,
\ee
in which case the quantization condition in \eqref{quantcond} becomes the well known Bethe ansatz equations
\be\label{heisBA}
\left(\frac{u_j+i/2}{u_j-i/2}\right)^{Q+J}=\prod_{k\ne j}^Q \frac{u_j-u_k+i}{u_j-u_k-i}\,.
\ee
The $u_j$ are sometimes referred to as ``Bethe roots''.
The energy of a particular state is additive with the individual energies of the magnons and has the form
\be
E=J+\sum_{j=1}^Q\,\eps_j\,
\ee
where 
\be
\eps_j=1+2g^2\frac{1}{u_j^2+1/4}\,.
\ee
The momenta $p_j$ must also satisfy one further condition that one does not normally encounter for the spin chain.  Since the gauge invariant operator has a trace, it is invariant under a single shift of  all the scalar fields inside the trace.  This translates into the  trace condition for the magnon momenta 
\be
\sum_{j=1}^Q\, p_j\,=\,0\mod 2\pi\,.
\ee

Now let us go into the thermodynamic limit and
assume $J \gg Q \gg 1$.  If we take the log on both sides of the Bethe equations in \eqref{heisBA} we have the approximate equation
\be\label{logbe}
i\, \frac{Q+J}{u_j}\approx 2i\,\sum_{k\ne j}\frac{1}{u_j-u_k}+2\pi\, n_j\, i\,,
\ee
where the integer $n_j$ arises from choosing a branch of the log.
If we define the density
\be
\rho(x)\equiv\sum_k\delta(x-u_k)
\ee
then the equation in \eqref{logbe} becomes an integral equation \cite{Minahan:2002ve,Beisert:2003xu}
\be\label{inteq}
2\pint \frac {dx'\rho(x')}{x-x'}={\frac{Q+J}{x}}\,-\,2\pi\, n_j\qquad x\in {\C_j} 
\ee
where the contour $\C_j$ refers to the contour where all roots are on the same log branch $n_j$.
Clearly the density must satisfy
\be\label{Qeq}
\int dx'\rho(x')=Q=\frac{{(J+Q)}-J+Q}{2}
\ee
where the second expression is given for later comparison with the $\sigma$-model.

In the thermodynamic limit we can approximate the momenta of the individual magnons as
\be
p_j\simeq 1/u_j
\ee
Hence, it follows that the total momentum $P$ is given by
\be\label{Peq}
P =\sum_j p_j =\int \frac{dx'\rho(x')}{x'}=P=0\mod 2\pi\,
\ee
where the last equality follows from the trace condition.  Furthermore the energy of each magnon may be approximated as 
\be
\epsilon_j\simeq1+2g^2/u_j^2\,,
\ee
from which one finds the total energy
\be
E=J+Q+2g^2\,\int \frac{dx'\rho(x')}{{x'}^2}\,,
\ee
We will rewrite this equation as 
\be\label{Eeq}
\int \frac{dx'\rho(x')}{{x'}^2}=\frac{1}{g^2}\,\frac{E-J-Q}{2}\,.
\ee

Hence, our task is to find solutions for $\rho(x)$ in the integral equations in  \eqref{inteq}, while satisfying the conditions in \eqref{Qeq} and \eqref{Peq}.  The integral equation was found by taking the classical limit $J,Q\to\infty$.  As such, the solutions should correspond to solutions of some classical system.  Indeed the classical system is the Landau-Lifschitz model which has the equation of motion
\be
\frac{\d}{\d t} \vec{S}=-2g^2\,\vec S\times \frac{\d^2}{\d \sigma^2} \vec S\,,
\ee
where $\sigma$ is the one dimensional position.
The spin $\vec S$ should be thought of as a continuous variable with fixed length $\half$.

Let us now turn to the general solutions of \eqref{inteq}.  We start by defining the resolvent $G(x)$,
\be\label{resolve}
G(x)\equiv\int\frac {dx'\rho(x')}{x-x'}
\ee
where the integral is over $N$ contours where $\rho(x')$ has support and $x$ is assumed to be off of the contours. We will assume that $\rho(x')$ is zero at the end points of the contours and so the resolvent will have square root branch points at these endpoints, with the branch cut giving a discontinuity across the contour.   A typical example  of the cuts is shown in figure 1.   Note that the cuts are symmetric under complex conjugation which  is a consequence of the Bethe equations and guarantees real values for the total energy.
\begin{figure}
  \centering
  \includegraphics{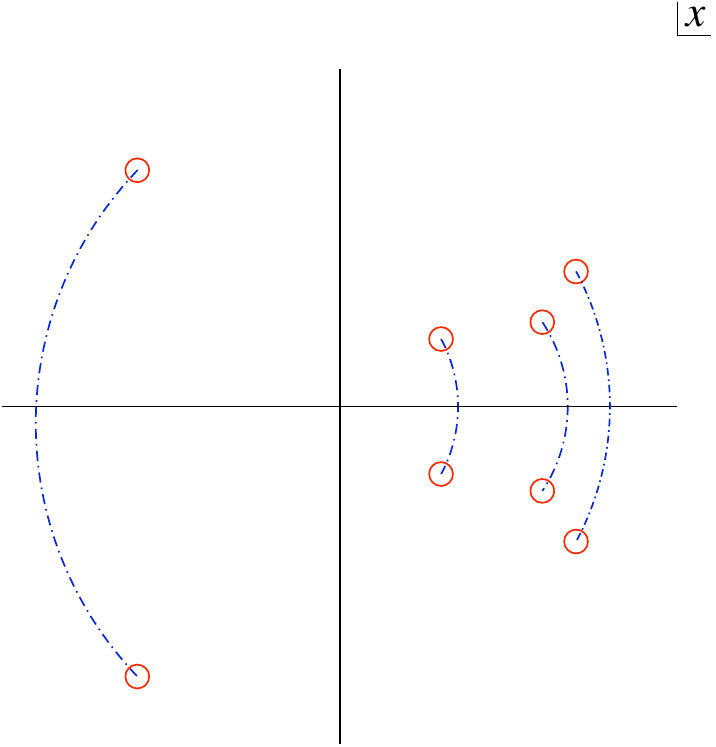}
   \caption{A solution with 4 branch cuts.  The different cuts correspond to different log branches and are symmetric under complex conjugation.}
  \label{fig:cuts}
\end{figure}

We can then extend the resolvent $G(x)$ to a two-sheeted surface connected by the $N$ branch cuts.  Hence this surface has genus $N-1$ and is hyperelliptic.   The top sheet is called the ``physical'' sheet and from \eqref{resolve} it is clear that $G(x)$ is nonsingular and single valued on this sheet.    Using the integral equation \eqref{inteq} we see that  along a contour $G(x_j)$ satisfies the condition
\be\label{Gcut}
G(x_j+i\eps)+G(x_j-i\eps)=\frac{Q+J}{x_j}\,-\,2\pi\,n_j\qquad x_j\in \C_j
\ee
Hence, we should expect $G(x)$ to have a pole on the bottom sheet at $x=0$.
We can also define two types of cycles, \A- and \B-cycles, where \A-cycles circle around single cuts on the top sheet and \B-cycles cross one cut onto the bottom sheet and cross another cut back onto the top sheet.  Such cycles are shown in figure 2.  If we integrate $dG$ around an \A-cycle, we find
\be
\oint_A dG=0
\ee
because of the single-valuedness of $G$.  On the other hand, integrating $dG$ around a \B-cycle that traverses cuts $j$ and $k$ gives
\be\label{Bcycle}
\oint_B dG=2\pi(n_j-n_k)
\ee
which follows from \eqref{Gcut}.
\begin{figure}
  \centering
  \includegraphics{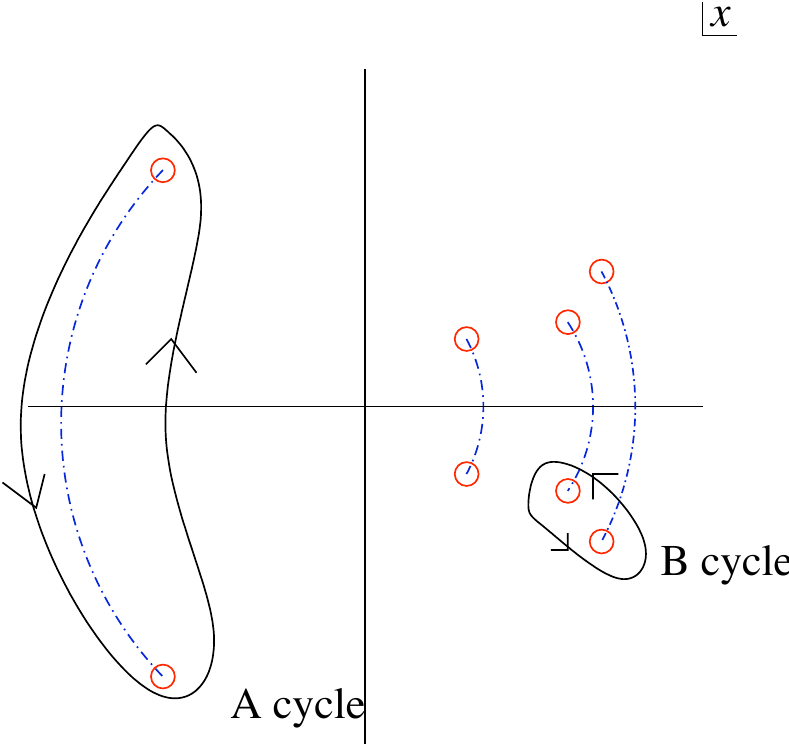}
 \caption{An \A-cycle circles around a single branch cut, while a \B-cycle traverses two cuts in going from the top sheet to the bottom sheet and back again.}
  \label{fig:cuts-AB}
\end{figure}

Now it is possible to recut the surface such that the hypersurface is unchanged.  This leaves the branch points where they are but changes how the branch points are paired up onto branch cuts.  These transformations correspond to $Sp(N-1,Z)$ transformations of the hypersurface.   These transformations should leave the integrals around the \A\ and \B-cycles unchanged, so this forces us to introduce condensates.   A condensate extends along the imaginary direction with a constant density which is quantized, namely $\rho(x)=-i n$ where $n$ is a positive integer.  The factor of $-i$ is so that $\int dx\rho(x)$ is positive real.  In figure 3 we recut the surface in figure 1 so that two branch cuts are redrawn.  A condensate now extends between the upper and lower branch cuts so that the \B-cycle remains nontrivial.  If the \B-cycle satisfies \eqref{Bcycle} then the condensate density is
\be
\rho(x)=-i(n_j-n_k)  \,.
\ee
In order that the original \A-cycles around the $j$ and $k$ cuts still give zero, we further require that the 
resolvent satisfy
\be
G(x+i\eps)+G(x-i\eps)=\frac{Q+J}{x}\,-\,2\pi\,n\qquad x\in \C
\ee
where $\C$ is either of the new contours and $n$ is some integer.  Note that $n$ has to be the same on both contours so that the \A-cycle integrals are zero.  
\begin{figure}
  \centering
  \includegraphics{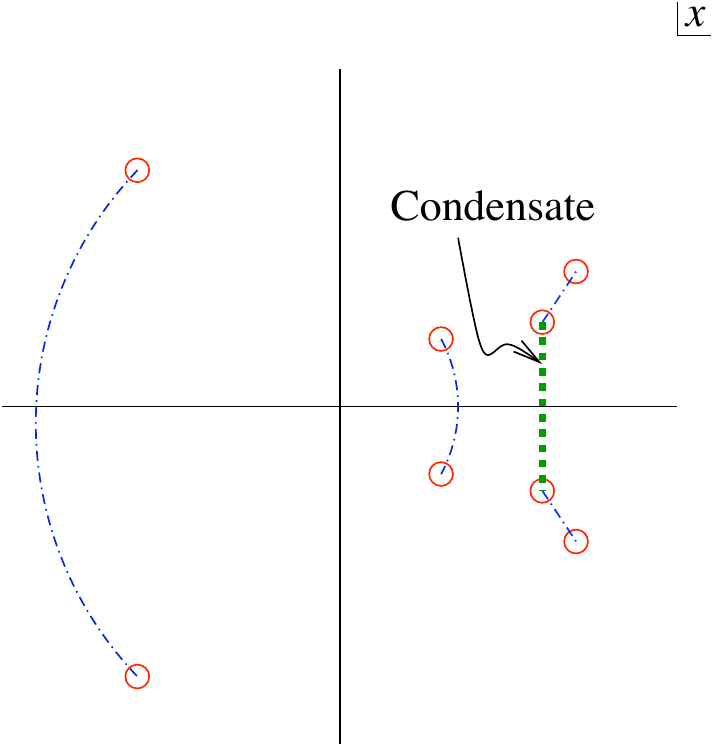}
   \caption{Recutting the hyperelliptic surface.  The condensate is now added to preserve the cycles.}
  \label{fig:recut}
\end{figure}
There is some freedom to move the condensate along the new branch cuts, but for simplicity we will place its end points on two of the branch points.  In order that there be no new poles  in the resolvent, it is necessary that the contour density at these branch points equals the condensate density.

\begin{figure}
  \centering
  \includegraphics{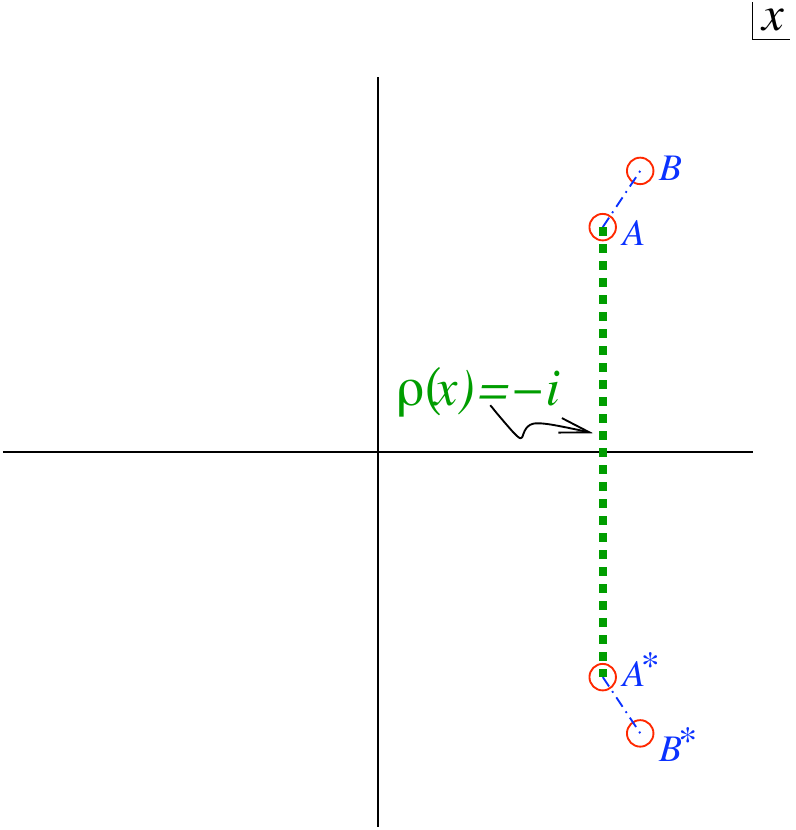}
   \caption{Two cuts joined by a condensate with density $\rho(x)=-i$.}
  \label{fig:single-magnon}
\end{figure}

Suppose for now that there are two contours and   recut the surface so that there is a condensate.  Let us further assume that the density of the condensate is $\rho(x)=-i$.  In general this configuration will not satisfy the trace condition, but this is certainly not a problem for the Heisenberg spin chain.  The resulting contours are shown in \figref{fig:single-magnon} where the branch points are labeled by $A$, $B$ and their complex conjugates.  

Let us now take the limit  $J\to\infty$.  In this limit the branch points $A$ and $B$ approach each other and the surface degenerates \cite{Vicedo:2007rp}.  In this limit we are left only with the condensate \cite{Minahan:2006bd,Vicedo:2007rp}.  It is then straightforward to show that 
\begin{eqnarray}\label{soliton}
Q&=&\int dx \rho(x)=-i\,(A-A^*)\nn\\
p&=&\int \frac{dx \rho(x)}{x}=-i\log(A/A^*)\nn\\
 E-J-Q&=&2g^2 \int \frac{dx \rho(x)}{x^2}=+2i\,g^2\,(A^{-1}-{A^*}^{-1})\,,
 \end{eqnarray}
from which it follows
\begin{eqnarray}
A&=&\frac{Q}{2}e^{ip/2}\csc\sfrac{p}{2}\nn\\
\label{hmag}\ve\equiv E-J&=&Q+8\,\frac{g^2}{Q}\,\sin^2\sfrac{p}{2}
\approx{\sqrt{Q^2+16g^2\sin^2\sfrac{p}{2}}}
\end{eqnarray}
where we have included the approximate expression for $E-J$ for comparison with the $\sigma$-model.  The expression $\ve$ corresponds to the energy of $Q$ bound magnons with total momentum $p$.   Note that if we had chosen the condensate density to be $\rho(x)=-i\,n$, then the resulting solution is $n$ sets of bound magnons, each with charge $Q/n$ and each with momentum $p$ 

The  classical solution that \eqref{soliton} corresponds to is  the soliton wave discussed by Lakshmanan, Takhtajan and Fogedby in the classical one-dimensional ferromagnet \cite{Lakshmanan:1977aa,Takhtajan:1977aa,Fogedby:1980aa}.  The soliton is a localized spin wave with a width and velocity dependent on the momentum.  The polar direction of the spin as a function of time and position $\sigma$ is given by
\be
\theta=2\,\sin^{-1}\left[\sin\sfrac{p}{2}\sech\frac{\sigma-vt}{\Gamma}\right]
\ee
where the width $\Gamma$ and the velocity $v$ are
\be
\Gamma=\frac{Q}{2\sin^2\sfrac{p}{2}}\qquad v=\frac{4g^2}{Q}\sin p
\ee

Let us now generalize the previous results for the Heisenberg spin chain to that of the $\sigma$-model on $R\times S^3$ with Virasoro constraints.  The details of how the integral equations are derived can be found in \cite{Kazakov:2004qf}, so here we only present the results.  First, the generalization of the integral equation in \eqref{inteq} is
\be\label{inteqs}
\pint \frac {dx'\rho(x')}{x-x'}={\frac{E\, x}{x^2-g^2}}-2\pi n_j\qquad x\in \C_j\,, 
\ee
where we see that the difference between \eqref{inteq} and \eqref{inteqs} is that the pole at $x=0$ has split into two poles at $x=\pm g$, and the residue has changed from $J+Q$ to $E$.  The only other change we need to make is that \eqref{Qeq} is modified to 
\be\label{Qeqs}
 \int dx'\rho(x')=\frac{{ E}-J+Q}{2}\,,
 \ee
 where like in \eqref{inteqs}, the difference between \eqref{inteqs} and \eqref{inteq} is that $J+Q$ is replaced with $E$.  The expressions \eqref{Peq} and \eqref{Eeq} are unchanged.
 
 If we now take the two cut solution, recut with a condensate, and then take $J\to\infty$, we are left only with the condensate \cite{Minahan:2006bd,Vicedo:2007rp}.  In this case we find
 \begin{eqnarray}
\sfrac12(E-J+Q)&=&-i(A-A^*)\nn\\
p&=&-i\log(A/A^*)\nn\\
 E-J-Q&=&\,+\,2i\,g^2\,(A^{-1}-{A^*}^{-1})\,,
 \end{eqnarray}
 from which we are lead to
\be\label{giant}
\ve \equiv E-J=\sqrt{Q^2+16g^2\sin^2\sfrac{p}{2}}
\ee

The classical solution that this condensate corresponds to  is the Hofman-Maldacena giant magnon \cite{Hofman:2006xt} if $Q=0$ or its generalization if $Q\ne0$ \cite{Chen:2006gea}.  Note that in the limit that $g\to0$ \eqref{giant} reduces to \eqref{hmag} for the bound magnons in the Heisenberg model.

\section{Finite size corrections for giant magnons}

In this section we find the leading order finite size corrections for the energy of a giant magnon  in the presence of an arbitrary number of other giant magnons.  With more than one magnon it is possible to satisfy the Virasoro constraint that the total momentum is $0$ mod $2\pi$, so the results will be gauge independent.

\begin{figure}
  \centering
  \includegraphics{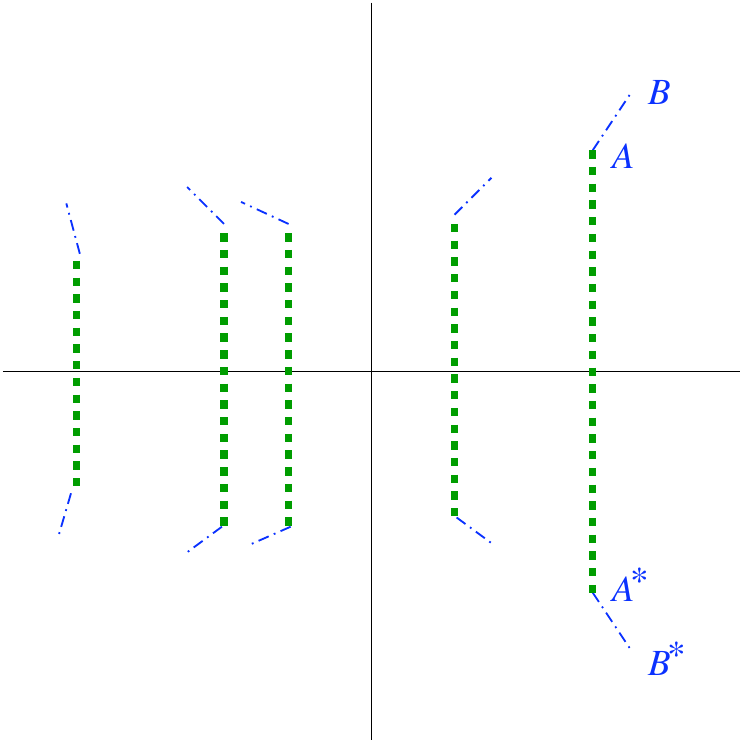}  
  \caption{Cuts for several finite size giant magnons}
  \label{fig:many-magnons}
\end{figure}

The root configuration corresponding to several finite size giant
magnons is depicted in \figref{fig:many-magnons}. A number of cuts are
pairwise connected by condensates. To calculate the dispersion
relation, we want to solve the equations \eqref{Peq}, \eqref{Eeq},
\eqref{Qeqs} and \eqref{inteqs}. Even for a single magnon, the
solution of \eqref{inteqs}, which leads to the AFZ result, is given by
an elliptic function.  For several magnons hyperelliptic functions are
needed.  The problem is, however, significantly simplified if we
assume each cut to be very short compared to the condensates.

For $x$ on a given cut $AB$, we can write the integral equation \eqref{inteqs} as
\begin{multline}
\pint_A^B \frac{dx'\rho(x')}{x-x'} + 2\,i\,\ln\frac{x-A}{x-A^*} + 
\int_{B^*}^{A^*} \frac{dx'\rho(x')}{x-x'} + \int_{\rm others} \frac {dx'\rho(x')}{x-x'} = \\
\frac{E}{2}\left(\frac{1}{x-g} + \frac{1}{x+g}\right) - 2\pi n.
\end{multline}
Except for the first two terms on the left hand side, this expression
depends only weakly on $x$. Thus, to a reasonable approximation,
\begin{equation}
  \pint_A^B \frac{dx'\rho(x')}{x-x'} = -2\,i\,\ln(x-A) + C,
  \label{eq:int-eq-C}
\end{equation}
where $C$ is a constant. By evaluation at $x = A$, $C$ can be expressed as
\begin{equation}
  C \approx \frac{E}{2} \left(\frac{1}{A-g}+\frac{1}{A+g}\right)-2\pi n+2\,i\,\ln(A-A^*)-2\,i\,\sum_j\ln\frac{A-A_j}{A-A_j^*}.
  \label{eq:C-approx}
\end{equation}

The problem has now been reduced to an integral equation over a single
contour, which can be solved for $\rho$ using a finite inverse Hilbert
transform \cite{Muskhelishvili:2008aa}, 
\begin{equation}
\rho(x) = -\frac{\sqrt{(B - x)(x - A)}}{2} \pint_A^B \frac{dx'}{x - x'} \frac{-2i\ln(x' - A) + C}{\sqrt{(B - x')(x' - A)}}.
\end{equation}
The term that involves the constant will vanish, while the rest of the
integral can be calculated to give
\begin{equation}
  \rho(x) = -\frac{2 i}{\pi} \arccos \sqrt{\frac{B - A}{x - A}}.
\end{equation}
Note that $\rho(B)=0$ while $\rho(A)=-i$.  Hence $\rho(x)$ is continuous where the cut joins the condensate, so the resolvent \eqref{resolve} will not have an extra pole there.  Such poles should only occur in the limit $A\to B$, where the surface degenerates \cite{Vicedo:2007rp} \footnote{We thank N. Dorey and B. Vicedo for discussions on this point.}.

In the $J \to \infty$ limit, the end point of the condensate is at
\be
  A_0 \equiv x_0 + i a_0 = \frac{Q + \ve}{4} e^{\frac{ip}{2}} \csc \frac{p}{2}
\ee
For finite $J$ this end point is shifted to $A \equiv A_0 + \delta A \equiv x_0 + i\, a$ and
the end point of the cut is at $B \equiv x_0 + i\, b$.  Since the cut
is assumed to be small we write
\begin{equation}
  B - A = i(b - a) \equiv i\,\delta\, e^{i\phi},
\end{equation}
with $\delta \ll a$. Plugging $\rho(x)$ back into \eqref{eq:int-eq-C},
we can now perform the integral to calculate $C$, which will be of
order $\order(\delta)$. Comparing this to \eqref{eq:C-approx} we
express the shift $B - A$ as
\begin{equation}
  \delta \,e^{i\phi} = 8 \, e^{-\frac{i E}{4}\lrbrk{\frac{1}{A+g} + \frac{1}{A-g}} + i \pi n}  
  \prod_{j} \frac{A - A_j^*}{A - A_j}.
  \label{eq:delta-phi-def}
\end{equation}

For $J \to \infty$, the charges $p$ and $Q$ are given by
\begin{equation}
  p = -i\,\ln\frac{A_0}{A_0^*}, \qquad Q = -i (A_0 - A_0^*) - i g^2(A_0^{-1} - {A_0^*}^{-1}).
\end{equation}
For finite $J$ there are additional contributions from the change of
the integral over the condensate, due to the shift of the end point,
and from integrals along the cuts.  Requiring that the shifts of $p$
and $Q$ vanish to the leading order, leads to the equations
\begin{equation}
  \frac{\delta}{2} \lrbrk{\frac{e^{i\phi}}{A_0} - \frac{e^{-i\phi}}{A_0^*}} - i\lrbrk{\frac{\delta A}{A_0} - \frac{\delta A^*}{A_0^*}} = 0,
\end{equation}
\begin{equation}
  \delta \cos \phi - \frac{g^2 \delta}{2} \lrbrk{\frac{e^{i\phi}}{A_0^2} + \frac{e^{-i\phi}}{{A_0^*}^2}} 
  - i\lrbrk{\frac{A_0^2-g^2}{A_0^2}\delta A - \frac{{A_0^*}^2-g^2}{{A_0^*}^2}\delta A^*} = 0,
\end{equation}
which is solved by
\begin{equation}
  \delta A=-\frac{i}{2}\,\delta\, e^{i\phi}.
\end{equation}
We see that the end point of the condensate splits into a cut.  The
shift in the dispersion relation is given by
\begin{equation}
  \Delta\ve = \delta \cos \phi + \frac{g^2 \delta}{2} \lrbrk{\frac{e^{i\phi}}{A_0^2} + \frac{e^{-i\phi}}{{A_0^*}^2}} 
  - i\lrbrk{\frac{A_0^2+g^2}{A_0^2}\delta A - \frac{{A_0^*}^2+g^2}{{A_0^*}^2}\delta A^*} = 0.
\end{equation}

The first nonvanishing correction occurs at order $\delta^2$. This
correction is calculated in the same way --- the contributions to the
charges from the shifted contours are calculated and required to
vanish. The resulting expressions for the shift of $A$ is given by the
somewhat complicated expression
\begin{multline}
  \delta A =  - \frac{i\delta}{2} e^{i\phi}
   - \frac{\delta^2}{\ve(\ve + Q)^2} 
   \Biggsbrk{\frac{8g^2 + (\ve + Q)^2}{8} \sin(p-2\phi) \\
   - g^2 \frac{\sin(3p-2\phi) - 4\sin 2\phi + \sin(p+2\phi)}{4} 
   + 4ig^2 \cos(p-2\phi) \sin^4\frac{p}{2}} + \order(\delta^3).
\end{multline}
The quadratic order correction to the dispersion relation, however, simplifies to
\begin{equation}
  \Delta\ve = -\frac{g^2}{a_0^2\,\ve} \sin^4\frac{p}{2} \,\cos(p-2\phi) \,\delta^2.
  \label{eq:disp-corr-gen}
\end{equation}
We will now consider this expression in several cases.

\subsection{Single fundamental magnon ($Q=0$)}

First we consider a single fundamental magnon with momentum $p$. In general such a
configuration will not satisfy the momentum condition. The expansion
parameters $\delta$ and $\phi$ are given by
\begin{equation}
  \delta\,e^{i\phi}=8a_0e^{-i\pi n}\exp(-E/\ve)=8a_0e^{-i\pi n}\exp(-(J+\ve)/\ve).
\end{equation}
Insertion into \eqref{eq:disp-corr-gen} gives
\begin{align}
  \Delta\ve &= -\frac{64 g^2}{e^2\,\ve} \sin^4\frac{p}{2} \,\cos(p + 2\pi n) \,\exp(-2J/\ve) \\
  &= -\frac{16 g\sin^3\frac{p}{2}}{e^2} \,{\cos(p + 2\pi n)} \,\exp\lrbrk{-2\frac{J}{4g\sin\frac{p}{2}}}.
\end{align}

Comparing this to the AFZ result we see that, for integer $n$, the results disagree by a factor of $\cos p$.
Of course, if we allow $n$ to be  noninteger then the $\cos{p}$ factor can be removed.  In fact this noninteger $n$ is a remnant of the missing momentum necessary to satisfy the level matching condition.  We will see shortly that with additional magnons, their momenta will enter into the argument of the cosine.

\subsection{Single dyonic magnon ($Q \ne 0$)}

For non-zero $Q$ the parameter $\delta$ and the phase $\phi$ are changed to
\begin{equation}\label{Qne0}
  \delta = 8\,a_0\,\exp(-E\,\ve/\ve_s^2), \qquad
  \phi = E\ \frac{Q \cot \frac{p}{2}}{\ve_s^2} - \pi n,
\end{equation}
with
\begin{equation}
  \ve_s^2\equiv\frac{Q^2}{\sin^2\frac{p}{2}}+16g^2\sin^2\frac{p}{2}.
\end{equation}
This results in a shift of the dispersion relation of the form
\begin{equation}
  \Delta\ve = -64 \sin^4\frac{p}{2} \,\cos(p - 2\phi) \frac{g^2}{e^{2\alpha}\,\ve} \exp(-2J\ve/\ve_s^2),
  \label{eq:disp-rel-shift-Q-nz}
\end{equation}
where
\begin{equation}
  \alpha = \frac{\ve^2}{\ve_s^2} = \frac{Q^2 \sin^2\frac{p}{2} + 16g^2\sin^4\frac{p}{2}}{Q^2 + 16g^2\sin^4\frac{p}{2}}.
\end{equation}
Note that for $E \gg \ve_s$ this correction is highly oscillatory.

Further note that we can get rid of the $e^{-2\alpha}$ term if we replace $ J$ by $ E$.  The interpretation of this is clear.  The size of a magnon is basically its energy, so we would expect the total size to be $J+\ve=E$.   Hence the exponential suppression factor should be governed by the size $E$ and not $J$.

\subsection{ Several fundamental magnons (all with $ Q=0$)}

Equation \eqref{eq:delta-phi-def} now reads
\begin{equation}
\delta_j \,e^{i\phi_j} = 8a_{0j} \prod_{k\ne j} \frac{A_j-A^*_k}{A_j-A_k} e^{-i\pi n}\exp(-E/\ve_j)\,,\qquad \ve_j=4g\sin\frac{p_j}{2},
\end{equation}
where our notation is extended in an obvious way to allow for several
parameters $\delta$, $\phi$, $a_0$ and $p$.  Using the lowest order
solution for $A_k$ and assuming $0 \le p_k < 2\pi$ we have
\begin{equation}
\prod_{k\ne j}\frac{A_j - A^*_k}{A_j - A_k} = \exp\lrbrk{-i\sum_{k\ne j}\frac{p_k}{2}} \frac{\sin\frac{p_j+p_k}{4}}{\sin\frac{p_j-p_k}{4}},
\end{equation}
resulting in
\begin{equation}\label{manyresult}
  \Delta\ve_j = -16\,g\,\sin^3\sfrac{p_j}{2}
  \prod_{k\ne j}^M\frac{\sin^2\frac{p_j+p_k}{4}}{\sin^2\frac{p_j-p_k}{4}}\,\exp(-2(J+\sum_k\ve_k)/\ve_j).
\end{equation}

Let us note a few points about the result in \eqref{manyresult}.  
\begin{itemize}
\item For finite $j$ the momenta will be quantized and this quantization will depend on the magnon \SM-matrix.  Hence \eqref{manyresult} applies for the $p_j$ that satisfy this quantization.  
\item Magnons come in flavors, but the result in \eqref{manyresult} is  flavor independent at this level.  In fact the flavors arise because of fermion zeromodes \cite{Minahan:2007gf}, which is obviously a quantum effect so one should not expect to see their effect in a classical result.  In a fully quantized theory we would therefore expect the flavor effects to be subleading in $1/g$.
\item The result in \eqref{manyresult} is singular if $p_j\to p_k$.  In this limit we are left with two magnons with identical momenta.  We will see the resolution of this in the next subsection.
\end{itemize}

\subsection{ $M$ magnons with the same momentum and $Q$}
To incorporate several magnons with the same momentum and spin, we can
make a small modification to the  integral equations.   Multiple magnons give multiple copies of the densities $\rho(x)$.  Hence the integral equations now read
\begin{equation}
M \pint_A^B \frac{dx' \rho(x')}{x - x'} = -2\,i\,M,\ln(x-A) + C_{ M}, 
\end{equation}
where $\rho(x)$ is the density for one magnon and
\begin{equation}
 C_M = \frac{E}{2} \lrbrk{\frac{1}{A - g} + \frac{1}{A + g}} - 2\pi n + 2M\,i\,\ln(A-A^*).
\end{equation}
The solution is the same as in the single magnon case, provided we make the substitutions
\begin{equation}
E\to \frac{E}{M}\qquad J\to\frac{J}{M}\qquad \phi\to \frac{\phi}{M}.
\end{equation}
The resulting shift of the dispersion relation is
\begin{equation}
\Delta\ve=-64\sin^4\frac{p}{2}\cos(p-2\frac{\phi}{ M})\frac{g^2}{e^{2\alpha}\,\ve}\exp(-2\frac{J}{ M}\ve/\ve_s^2).
\end{equation}

This result is more transparent in the $Q \to 0$ limit, where it simplifies to
\begin{equation}
\Delta \ve = -\,\frac{16 g\sin^3\frac{p}{2}}{e^2}\, 
  \cos(p + 2\pi \frac{n}{M})\,\exp\lrbrk{-2 \frac{J/M}{4g \sin\frac{p}{2}}}.
\end{equation}
A physical string would have $p=2\pi\, m/M$, for some integer $m$.
Hence we can remove the $\cos$--factor by a suitable choce of $n$.
This then reproduces AFZ's result for these special momenta, as well
as the orbifold results of Astolfi \textit{et al.} \cite{Astolfi:2007uz}.  The physical strings with these $M$ multiple magnons correspond to the limit of the GKP spinning string on $S_5$ \cite {Gubser:2002tv} for $M=2$, or their multicusp generalizations  \cite{Kruczenski:2004wg} when $Q=0$.  If $Q\ne0$, then the physical strings are the helical strings described in \cite{Okamura:2006zv}.

Note that the exponential suppression is not as large as for single magnons.  Looking at \eqref{manyresult}, we see that the singularity as two magnons approach the same momentum is signifying this change in the suppression factor.

\section{Checking results --- back to Heisenberg}

For $Q \gg g$ we expect the string result to approach the result
from the one loop gauge theory calculation,
\textit{i.e.}~from the Heisenberg model. For infinite spin $J$ we have
\begin{equation}
\ve\simeq Q+\frac{8g^2}{Q}\sin^2\frac{p}{2}.
\end{equation}
From \eqref{eq:disp-rel-shift-Q-nz} the finite size correction in this limit is given by
\begin{align}
\Delta\ve & = -64 \sin^4\frac{p}{2} \cos(p-2\phi) \frac{g^2}{\ve} \exp\lrbrk{-2\ve\frac{\ve+J}{\ve_s^2}} \\
     & \simeq -64 \sin^4\frac{p}{2} \cos(p-2\phi) \frac{g^2}{Q}     \exp\lrbrk{-2\frac{J+Q}{Q} \sin^2\frac{p}{2}},
     \label{eq:dE-large-Q}
\end{align}
where the phase is given by
\begin{equation}
  \phi = E\ \frac{Q\cot\frac{ p}{2}}{\ve_s^2}-\pi n\approx (J+Q)\ \frac{\sin p}{2Q}-\pi n.
\end{equation}

It is instructive to  reproduce \eqref{eq:dE-large-Q} from the Heisenberg
model.
We start by quickly recalling the ``Bethe string'' solutions of the Bethe equations in the infinite $J$ limit.
 Consider again the Bethe equation
\begin{equation}
  \lrbrk{\frac{u_j+i/2}{u_j-i/2}}^{Q+J} = \prod_{k\ne j}^Q \frac{u_j-u_k+i}{u_j-u_k-i}.
  \label{heisBA-again}
\end{equation}
Assume that $\Im u_j > 0$ for some $j$. As $J\to\infty$ the left hand
side will grow indefinitely. For \eqref{heisBA-again} to be satisfied, there
has to be a pole on the right hand side. We will have a pole if there is another root, $u_{j+1}$ such that $u_j-u_{j+1}\to i$.  If $\Im u_{j+1}>0$ then its Bethe equation will also have a divergent left hand side.  But its right hand side already has a zero since $u_{j+1}-u_{j}\to -i$.   Hence to compensate, there must be a $u_{j+2}$ with $u_{j+1}-u_{j+2}\to i$ such that $\frac{u_{j+1}-u_{j+2}-i}{u_{j+1}-u_{j}+i}\to 0$.  If we continue finding roots in this way, eventually we will have a root $u_k$ where $\Im u_k<0$.  In this case, its Bethe equation will approach $0$ on the left hand side, so the right hand side will also have to be zero, hence we would need $u_k-u_{k+1}\to i$ such that $\frac{u_k-u_{k-1}+i}{u_k-u_{k+1}-i}\to0$, and so the argument works in reverse from the $\Im u_j>0$ case.  

\begin{figure}
   \centering
   \includegraphics[height=6cm]{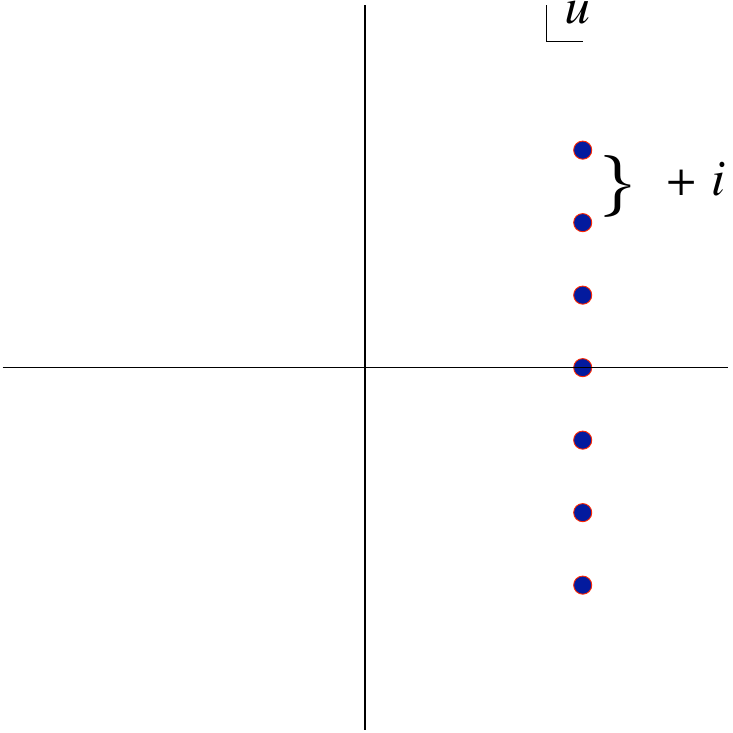}  
   \caption{Bethe string}
   \label{fig:bethestring2}
 \end{figure}

Thus, as $J\to\infty$ the $Q$ Bethe roots will
distribute themselves along a line parallel to the imaginary
axis, each root separated by $i$. For the energy to be real, the configuration has to be symmetric
under complex conjugation. The top root will then be $u_1 = x_0 +
i(Q-1)/2$ and the bottom root $u_Q = x_0 - i(Q-1)/2$.  This distribution is shown in \figref{fig:bethestring2}.

To calculate the momentum $p$ we use the equation
\begin{equation}
  e^{ip} = \prod_j \frac{u_j + i/2}{u_j - i/2} = \frac{u_1 + i/2}{u_Q - i/2} = \frac{x_0 + \frac{iQ}{2}}{x_0 - \frac{iQ}{2}},
\end{equation}
which we can solve for the parameter $x_0$ to get
\begin{equation}
  x_0 = \frac{Q}{2} \cot\frac{p}{2}.
\end{equation}
The dispersion relation is given by
\begin{equation}
  \ve = Q + i g^2 \sum_{j=1}^Q \lrbrk{\frac{1}{u_j + i/2} - \frac{1}{u_j - i/2}} = Q + ig^2\lrbrk{\frac{1}{u_1+i/2} - \frac{i}{u_Q-i/2}} = 
  Q + \frac{8g^2 Q^2}{Q^2 + 4 x_0^2}.
\end{equation}
Using the above solution for $x_0$ we get
\begin{equation}
  \ve = Q +  \frac{8g^2}{Q} \sin^2 \frac{p}{2}
\end{equation}
as expected.

If we now consider 
$J$ to be large but finite, the roots will not be exactly separated by $i$.  To see how the distribution changes, suppose that $u_1$ is the top root in \figref{fig:bethestring2} and let $u_1-u_2=i+\eps$.  From the above argument we have that $\frac{u_2-u_3}{u_1-u_2}\to 0$ as $J\to \infty$, hence we expect
  $u_2 - u_3 = i + \order(\eps^2)$,   $u_3 - u_4 = i + \order(\eps^3)$, {\it etc.}, until we reach a $u_j$ that crosses the real axis.  Once this happens the degree of separation reverses symmetrically. Hence the
leading order finite size corrections will be governed by the
small change in the positions of the top and bottom roots.  This is similar to what we
had for the finite gap equations, where the corrections corresponded
to additional cuts at the ends of the condensates.

Let us apply this to the case of a single set of bound magnons.
  Writing out the Bethe
equation for $u_1 = x_0 + i(Q-1)/2$ we get
\begin{equation}
\left(\frac{x_0+i\,Q}{x_0+i(Q-1)}\right)^L \simeq \frac{2\cdot3\dots (Q-1)\cdot Q}{\eps\cdot 1\cdot 2\dots(Q-2)} \simeq \frac{Q^2}{\eps},
\end{equation}
where $L \equiv Q + J$.  For large $L$, the left hand side approaches
$\exp\lrbrk{\frac{iL}{x_0+iQ}}$, and the parameter $\eps$ can be calculated as
\begin{equation}
\eps=Q^2e^{-2L\sin^2\frac{p}{2}/Q}e^{2i\phi},\qquad\phi=\frac{L}{2Q}\sin p.
\end{equation}
Note that when comparing to \eqref{Qne0}, $\eps\sim \delta^2e^{2i\phi}$. Hence we expect nonvanishing
finite size corrections already at leading order for the Heisenberg
model.

For finite $L$, $x_0$ will be shifted to $x$ and the momentum will be
shifted by
\begin{equation}
e^{i\Delta p}=\frac{x+iQ/2+i\eps}{x+iQ/2}\cdot\frac{x-iQ/2}{x-iQ/2-i\eps^*}.
\end{equation}
The shift in the dispersion relation can be calculated as
\begin{equation}
\Delta \ve = 2g^2 \lrbrk{ \frac{i}{x+iQ/2+i\eps} - \frac{i}{x+iQ/2}+ \text{ c.c. } }.
\end{equation}
By adjusting $x$ so that $\Delta p=0$, we finally can express the
correction to the dispersion relation as
\begin{equation}
\Delta\ve = -64 \sin^4\frac{p}{2} \cos(p-2\phi) \frac{g^2}{Q} \exp\lrbrk{-2\frac{J+Q}{Q} \sin^2\frac{p}{2}}.
\end{equation}
This result from the Heisenberg model perfectly agrees with
\eqref{eq:dE-large-Q} in the $Q \gg g$ limit of the finite gap
solution.
This procedure can
be generalized to several sets of magnons, where one will continue to find agreement with the finite gap calculation.  However in the case of condensates with density $-i n$ we need to be careful, since in the Bethe ansatz, it is not possible to have a Bethe string with density greater than $-i $ \cite{Beisert:2005mq} \footnote{We thank K. Zarembo for comments on this point.}.  What we believe happens here is that there will be $n$ nearby strings, with a  small separation in the real direction between each string.

\section{ Results from TBA}

As a final check on our results, we consider the leading finite size corrections  using the Thermodynamic Bethe ansatz .  The TBA  was first applied to $\NN=4$ by  Ambjorn, Janik and Kristjansen \cite{Ambjorn:2005wa}, where they showed that it leads to corrections of order $g^{2L}$ in the weak coupling expansion, precisely the order at which wrapping effects should appear \cite{Serban:2004jf}.  More recently  Janik and {\L}ukowski applied the TBA to an explicit computation of  the finite size corrections of a single fundamental magnon \cite{Janik:2007wt}.  For another interesting application see \cite{Arutyunov:2007tc}.

Applying the TBA with more than one magnon would appear to be a  hard problem.  In order to find the complete leading order correction for $M$ magnons, including prefactors, one would need to compute \SM-matrices for $M+1$ particles \cite{Janik:2007wt}.  Although the final answer in \eqref{manyresult} is quite simple, we have not yet attempted this computation.   The \SM-matrices are flavor dependent, but we expect this dependence to drop out when computing the leading order contribution.

Instead we will focus on the more modest goal of reproducing our results for one dyonic giant magnon of charge  $Q$  and limit ourselves further by only finding the leading exponential term and not the prefactor \footnote{As this paper was being prepared \cite{Hatsuda:2008gd} appeared which does calculate the prefactor.}

\begin{figure}
  \centering
  \includegraphics{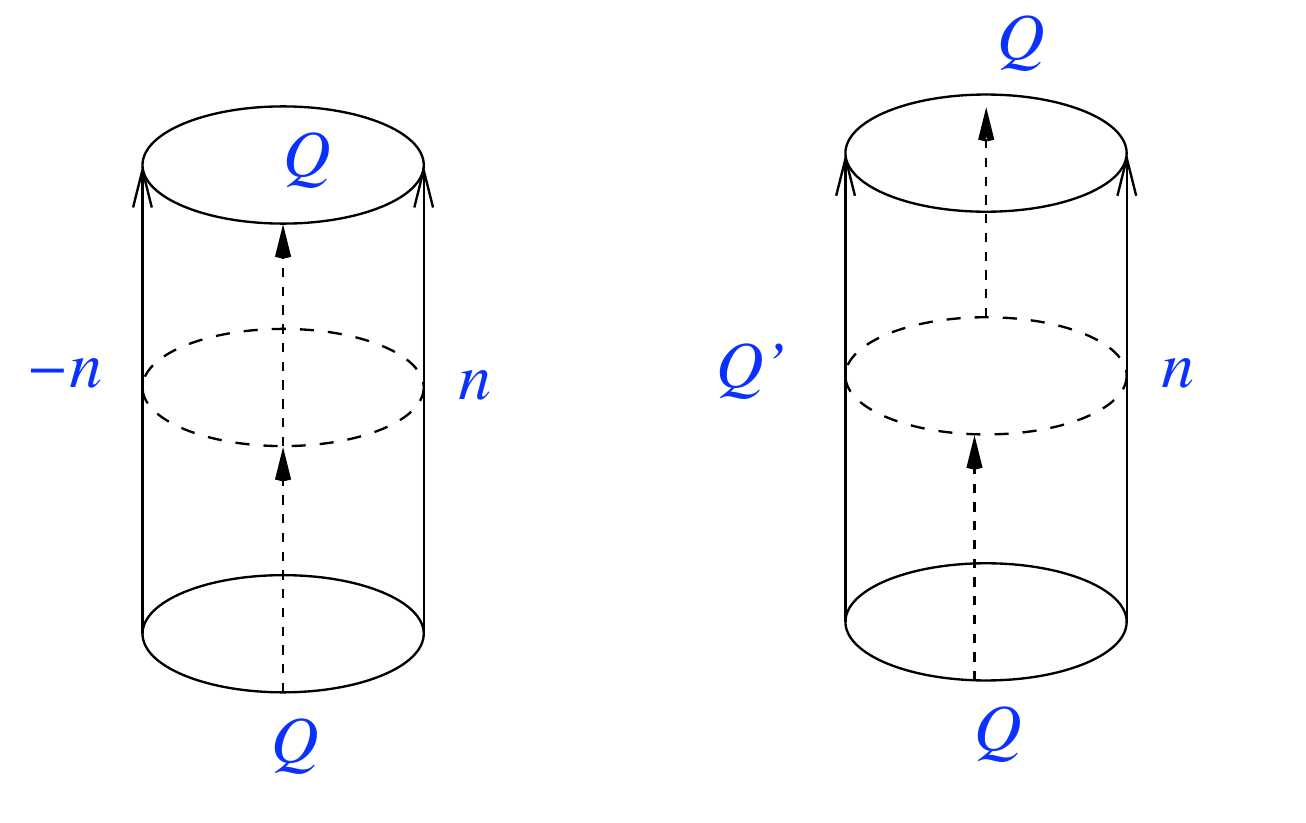}  
  \caption{World-sheet diagrams corresponding to the $F-$ and $\mu$-terms.  We have indicated the charge of the magnon as well as the virtual particle charges.}
  \label{fig:mu-f-term}
\end{figure}

As was argued in \cite{Ambjorn:2005wa,Janik:2007wt}, there can be two types of graphs that contribute to finite size effects, as was first shown for a relativistic field theory by L\"uscher \cite{Luscher:1985dn}.  These are called $F$- and $\mu$-terms and are shown in \figref{fig:mu-f-term}.  For the $F$-term  graph the particle emits and absorbs an on-shell virtual particle that travels around the compact dimension.  For the $\mu$-term graph, the particle splits into two virtual particles that travel in opposite directions around the compact direction and recombine into the original particle.  It turns out that the $\mu$-term is responsible for the finite size corrections to the giant magnon energy \cite{Janik:2007wt}.

The generalization of the $\mu$-term graph in \cite{Janik:2007wt} is the 
creation/annihilation of  two on-shell particles with charges $n$ and $Q'$, where $n\ge1$ and we assume $Q'\gg n$.  The charges are highest weights of  $SU(2|2)\times SU(2|2)$, and so in order to conserve this nonabelian charge, they must satisfy $Q-n\le Q'\le Q +n$.  Defining $Q'\equiv Q+n'$, we see that $-n\le n'\le n$. The exponentially suppressed correction then arises from the 
 $\exp({ip_cJ})$ factor that appears in the virtual particle propagator, where $ p_c$ is the virtual particle momentum.  The other virtual particle has momentum defined to be $-\delta p$, so
on shell conservation of momentum gives $p_c=p+\delta p$.  Requiring energy conservation then leads to
\begin{equation}\label{conseq}
\sqrt{Q^2+16g^2\sin^2\sfrac{p}{2}}=\sqrt{(Q+n')^2+16g^2\sin^2\sfrac{p+\delta p}{2}}+\sqrt{n^2+16g^2\sin^2\sfrac{\delta p}{2}}
\end{equation}

To leading order \eqref{conseq} reduces to 
\begin{equation}
\ve_s^2\delta p^2-4n'Qg^2\cot\sfrac{p}{2}\,\delta p+4n^2=0
\end{equation}
which has the solution
\begin{equation}
\delta p=2n'\frac{Q\cot\sfrac{p}{2}}{\ve_s^2}+\,i\,\frac{2}{\ve_s^2}\sqrt{n^2\ve_s^2-{n'}^2Q^2\cot^2\sfrac{p}{2}}
\end{equation}
We wish to minimize the imaginary part to lessen the exponential suppression.  Clearly this occurs when $n=1$ and $n'=\pm1$ and so we find
\be\label{deltap}
\delta p=\pm 2\frac{Q\cot\sfrac{p}{2}}{\ve_s^2}+\,2i\,\frac{\ve}{\ve_s^2}\,.
\ee
  Both possibilities for $n'$ will contribute to the $\mu$-term graphs.  
If we now substitute \eqref{deltap} into $\exp(ip_c J)$, we obtain the exponential suppression in \eqref{eq:disp-rel-shift-Q-nz} as well as the phase in \eqref{Qne0}.

\section{Discussion}

In this article we have derived the leading finite size corrections for a general configuration of giant magnons on physical strings, that is when the total momentum is a multiple of $2\pi$.   We have found that the form of the corrections is particularly simple if all magnons have $Q=0$.  We have also shown that while the finite gap configuration for a general collection of giant magnons involves hyperelliptic functions, the leading order computation is found by reducing to a single cut and hence ordinary functions.  

One can go on and find the next order corrections as well.  The method is straightforward, albeit tedious.  Again the computation can be reduced to single cut configurations and hence ordinary functions.  One might also want to compare these results with other methods, for example by examining multi-soliton solutions in the sine-Gordon model.

Another interesting question is to compute the quantization of the magnon momenta for large but finite $J$.  Moreover, it would be interesting to see how a particular set of magnons with quantized momenta connect onto string solutions with small values of $J$.
Finally, it should be possible to compute the leading order corrections using the methods of Janik and {\L}ukowski \cite{Janik:2007wt}.

\bigskip
\subsection*{Acknowledgments}
\bigskip
We thank N.~Dorey, V.~Forini, S.~Frolov, R.~Janik, T.~Klose, T~McLoughlin, G.~Semenoff, B.~Vicedo, M.~Zamaklar and K.~Zarembo for many helpful discussions.  We also thank K.~Zarembo for comments on the manuscript.  J.~A.~M. would like to thank the Newton Institute at Cambridge University and the CTP at MIT for hospitality during the course of this work.  This research was supported in part by Vetenskapr\aa det and the STINT foundation.

\bibliographystyle{nb}
\bibliography{finite_size}
 
\end{document}